\documentclass[twocolumn,showpac,showkeys,preprintnumbers,amsmath,amssymb]{revtex4}

\usepackage{graphicx}
\usepackage{epsfig}
\usepackage{wrapfig}
\usepackage{rotating}
\usepackage{color}
\usepackage{comment}
\usepackage[normalem]{ulem}
\usepackage{multirow}
\usepackage{tabularx}
\usepackage{graphicx}
\usepackage{dcolumn}
\usepackage{hyperref}
\usepackage{float}
\usepackage{amsmath}
\usepackage{amssymb}
\usepackage{lipsum}
\usepackage{graphics}

\usepackage{color}

\newcolumntype{R}[1]{>{\raggedleft\arraybackslash}p{#1}}

\begin{document}

\title{Simulations of neutrino and gamma-ray production from relativistic black-hole microquasar jets}
\author{Th. V. Papavasileiou$^{1,3}$}\email{th.papavasileiou@uowm.gr}
\author{O. T. Kosmas$^2$}\email{odysseas.kosmas@manchester.ac.uk}
\author{J. Sinatkas$^1$}\email{isinatkas@uowm.gr}
\affiliation{$^1$Department of Informatics, University of Western Macedonia, GR-52100 Kastoria, Greece}
\affiliation{$^2$Modelling and Simulation Center, MACE, University of Manchester, Sackville Street, Manchester, UK}
\affiliation{$^3$Division of Theoretical Physics, University of Ioannina, GR-45110 Ioannina, Greece}

\begin{abstract} 
 
\end{abstract}

\begin{color}{red}
\end{color}








  
\begin{abstract}
Recently, microquasar jets have aroused the interest of many researchers focusing on the astrophysical plasma outflows and various jet ejections. In this work, we concentrate on the investigation of electromagnetic radiation and particle emissions from the jets of stellar black hole binary systems characterized by their hadronic content in their jets. Such emissions are reliably described within the context of the relativistic magneto-hydrodynamics. Our model calculations are based on the Fermi acceleration mechanism through which the primary particles (mainly protons) of the jet are accelerated. As a result, a small portion of thermal protons of the jet acquire relativistic energies, through shock-waves generated into the jet plasma. From the inelastic collisions of fast (non-thermal) protons with the thermal (cold) ones, secondary charged and neutral particles (pions, kaons, muons, $\eta$-particles, etc.) are created as well as electromagnetic radiation from the radio wavelength band, to X-rays and even to very high energy $\gamma$-ray emission. One of our main goals is, through the appropriate solution of the transport equation and taking into account the various mechanisms that cause energy losses to the particles, to study the secondary particle distributions within hadronic astrophysical jets. After testing our method on the Galactic MQs SS 433 and Cyg X-1, as a concrete extragalactic binary system, we examine the LMC X-1 located in the Large Magellanic Cloud, a satellite galaxy of our Milky Way Galaxy. It is worth mentioning that, for the companion O star (and its extended nebula structure) of the LMC X-1 system, new observations using spectroscopic data from VLT/UVES have been published few years ago.
\end{abstract}

\maketitle


\section{Introduction}

In recent years, astrophysical magnetohydrodynamical flows in Galactic and extragalactic microquasars 
(MQs) and X-ray binary systems (XRBs) have been modelled with the purpose of studying their 
multi-messenger emissions (e.g. neutrinos, gamma-ray emissions) \cite{Romero,Reynoso, Rev_Romero}. For the detection of such emissions, extremely sensitive detector 
tools are in operation for recording their signals reaching the Earth
like IceCube, ANTARES, KM3NeT \cite{IceCube-PRL, IceCube, KM3Net_2016}, etc.
Modelling offers good support for future attempts to detect radiative multiwavelength emission and particles, 
while in parallel several numerical simulations have been performed towards 
this aim \cite{Ody-Smpon-2015,Ody-Smpon-2017,Ody-Smpon-2018}.

Microquasars (or generally XRBs) are binary systems consisting of a compact stellar object (usually a black hole or a neutron star) and a donor (companion) star \cite{Mirabel}. Mass from the companion star overflows through the system's Langrangian points to the equatorial region of the compact object gaining angular momentum and thus forming an accretion disc of very high temperature gas and matter. This mass expelling could be mainly due to a concentrated stellar wind \cite{Friend,Hanke,Hell}. Consecutively, a portion of the disc's matter is being concentrated by the system's magnetic field (initially it is attached to the rotating disc) and is ejected perpendicular to the disc in two opposite directions \cite{Falcke} forming the system's jets. The jets are detectable from the Earth even when the system's distance is too large. This is due to the relativistic velocities they acquire \cite{Aharonian2005, Albert2006,Albert2007} combined with Doppler effects when they are headed towards the Earth. It has been shown that the kinetic luminosity coming from astrophysical microquasar jets constitutes a substantially large part of the total galactic cosmic radiation \cite{Heinz}. Moreover, it has been proved that microquasar binary systems that don't produce thermal jets are more likely to be neutrino and gamma-ray emission sources \cite{Waxman}.

The most well-studied microquasar systems include the Galactic X-ray binaries SS433, Cyg X-1, Cyg X-3, etc. \cite{Romney, Reid}, while from the extragalactic systems we mention the LMC X-1, LMC X-3 (in the neighbouring galaxy of the Large Magellanic Cloud) \cite{diBenedetto, MSc-Papav}, and the Messier X-7 (in the Messier 33 galaxy)\cite{M33}. Their respective relativistic jets are emission sources in various wavelength 
bands and high energy neutrinos. In this work, we focus our study on the extragalactic binary system LMC X-1 with the purpose to determine its gamma-ray and neutrino emissions produced through the processes and mechanisms that are about to be discussed. Concerning the binary system LMC X-1, new studies and observational results have been recorded, shedding more light on its unique spectral and environmental characteristics \cite{Hyde, Cooke} and thus providing a more solid base for further in-depth analysis.  

Up to now, the SS433 is the only microquasar observed with a definite hadronic content 
in its jets, as verified from observations of their spectra (see Ref. 
\cite{Ody-Smpon-2015,Ody-Smpon-2017,Ody-Smpon-2018} and references therein). 
Radiative transfer calculations may be performed at every 
point in the jet for a range of frequencies (energies), at every location 
\cite{Smponias_tsk_2011} providing the relevant emission and absorption coefficients. 
Line-of-sight integration, afterwards, provides synthetic images of $\gamma$-ray emission, 
at the energy-window of interest \cite{Smponias_tsk_2011,Smponias_tsk_2014}. 

Studies of the concentrations of jet particles, whose interactions lead to neutrino and gamma-ray production, need to take into account various energy loss mechanisms that occur due to several hadronic processes, particle decays and particle scattering \cite{Romero,Reynoso}. In the known fluid approximation, macroscopically the jet matter behaves as a fluid collimated by the magnetic field. At a smaller scale, consideration of the kinematics of the jet plasma becomes necessary for treating shock acceleration effects.
 
In the model employed in this work \cite{Romero, Reynoso}, the jets are considered to be conic 
along the z-axis (ejection axis) with a radius $r(z) = z \tan{\xi}$, where $\xi$ 
its half-opening angle \cite{Marshall2002}. The jet radius at its base, $r_0$, is given by 
$r_0 = z_0 \tan{\xi}$, where $z_0$ is the distance of the jet's base to the 
central compact object. According to the jet-accretion speculation, only 10\% of 
the system's Eddington luminosity \cite{Kording} $(L_k=1.2\times 10^{37}M$ $erg/s$, M in solar 
masses$)$ is transferred to the 
jet for acceleration and collimation through the magnetic field given by the 
equipartition of magnetic and kinetic energy density as $B=\sqrt{8\pi\rho _k(z)}$
(see Ref. \cite{Romero,Reynoso,Smponias_tsk_2011,Smponias_tsk_2014}). 

By assuming the one-zone approximation \cite{Khangulyan}, we consider a small portion of the hadrons (mainly protons) equal to $q_r \approx 0.1$ to be accelerated in a zone from $z_0$ to $z_{max}$ with the rate $t_{acc}^{-1}\simeq\eta ceB/E_p$ \cite{Gallant} according to the 2nd order Fermi acceleration mechanism. The particles are accelerated to nearly relativistic velocities (with $\eta =0.1$
being the acceleration efficiency) \cite{Begelman} resulting in a power-law distribution 
given in the jet's rest frame by $N'(E')=K_0E'^{-2}$ [$GeV^{-1}cm^{-3}$],
where $K_0$ is a normalization constant.

In the rest of the paper, after a brief description (sec. II) of the p-p interaction chain leading to neutrino and gamma-ray emission, we solve the transfer equation (sec. III) to determine the energy distributions of the primary and secondary jet particles. Then (sec. IV), the obtained predictions for high-energy neutrino and gamma-ray production, for the LMC X-1 MQ system, are presented and discussed. Finally (sec. V), we summarize the main conclusions.   

\section{Interaction chain leading to neutrino and gamma-ray production}

In general, the main interactions of the relativistic protons include those with the stellar winds \cite{Romero2003,Romero2007}, the radiation fields (composed of internal and external emission sources) \cite{Vila} as well as the cold hadronic matter of the jet. In this work, we focus on the last interaction because it is more dominant. 

Neutrino production and gamma-ray emission are the results of a reaction chain that is caused by the p-p (and p-$\gamma$) interactions inside the jet. KM3NeT, ANTARES and IceCube \cite{IceCube-PRL, IceCube, KM3Net_2016} are prominent examples of undersea water and under-ice detectors that are able to detect those neutrinos that reach the Earth. The aforementioned reaction chain begins with inelastic p-p collisions 
of the relativistic protons with the cold ones inside the jet, which generate neutral 
($\pi^0$) and charged ($\pi^\pm$) pions according to the following reactions
\begin{align}
pp&\rightarrow pp + \alpha\pi ^0 + \beta\pi ^{+}\pi ^{-} \nonumber\\
pp&\rightarrow pn + \pi ^{+} + \alpha\pi ^0 + \beta\pi ^{+}\pi ^{-} \nonumber\\ 
pp&\rightarrow nn + 2\pi^{+} + \alpha\pi ^0 + \beta\pi ^{+}\pi ^{-} \nonumber
\end{align}
with $\alpha$ and $\beta$ being the particle plurality factors that depend on the proton energy \cite{Mannheim}. Neutral pions decay into gamma-ray photons as
\begin{equation}
\pi ^0\rightarrow\gamma +\gamma
\end{equation}
while the charged pions decay into muons and neutrinos as
\begin{equation}
\pi^{+}\rightarrow\mu ^{+}\nu_{\mu}\rightarrow e^{+}\bar{\nu}_{\mu}\nu_{e}+\nu_{\mu} \nonumber
\end{equation}
\begin{equation}
\pi^{-}\rightarrow\mu^{-}\bar{\nu}_{\mu}\rightarrow e^{-}\bar{\nu}_{e}\nu_{\mu}+\bar{\nu}_{\mu} \nonumber
\end{equation}
Subsequently, muons also decay into neutrinos.
These are the main reactions feeding the neutrino and gamma-ray production channel in the 
employed model.

\begin{figure*}[ht] 
\centering
\includegraphics[width=.45\linewidth]{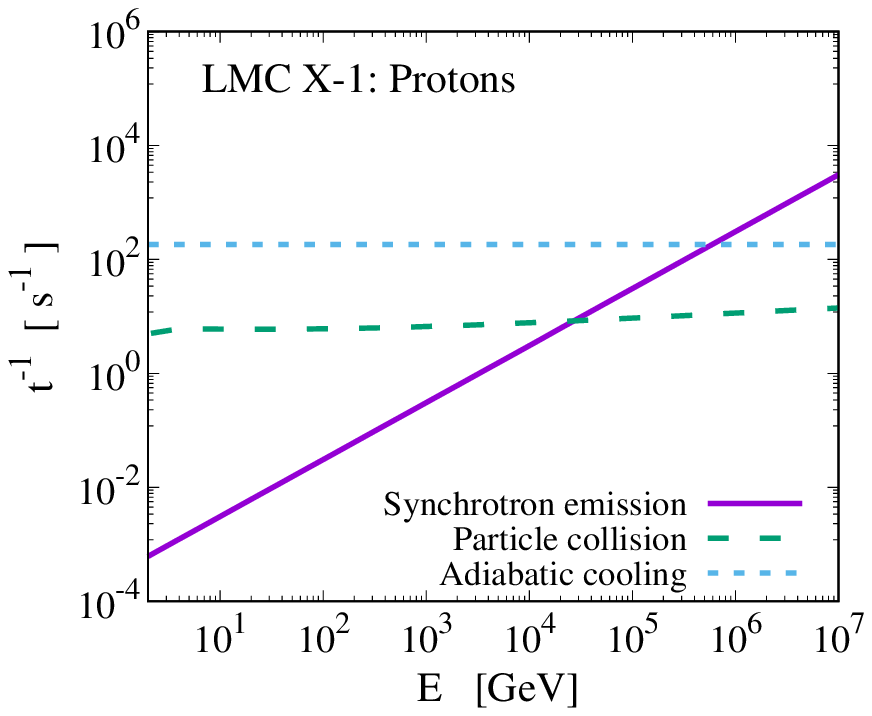}
\hspace*{-0.6 cm}
\includegraphics[width=.45\linewidth]{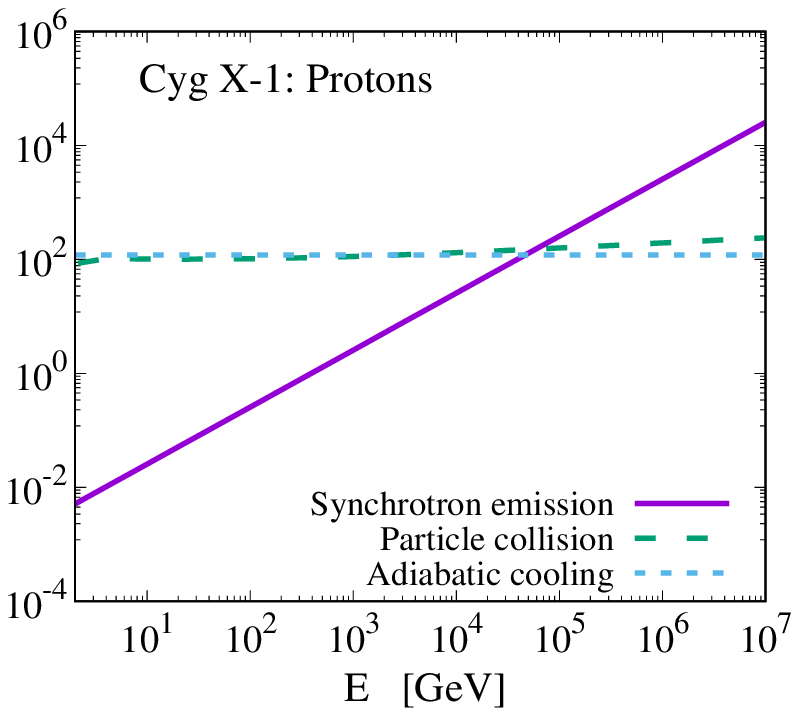}\\
\includegraphics[width=.45\linewidth]{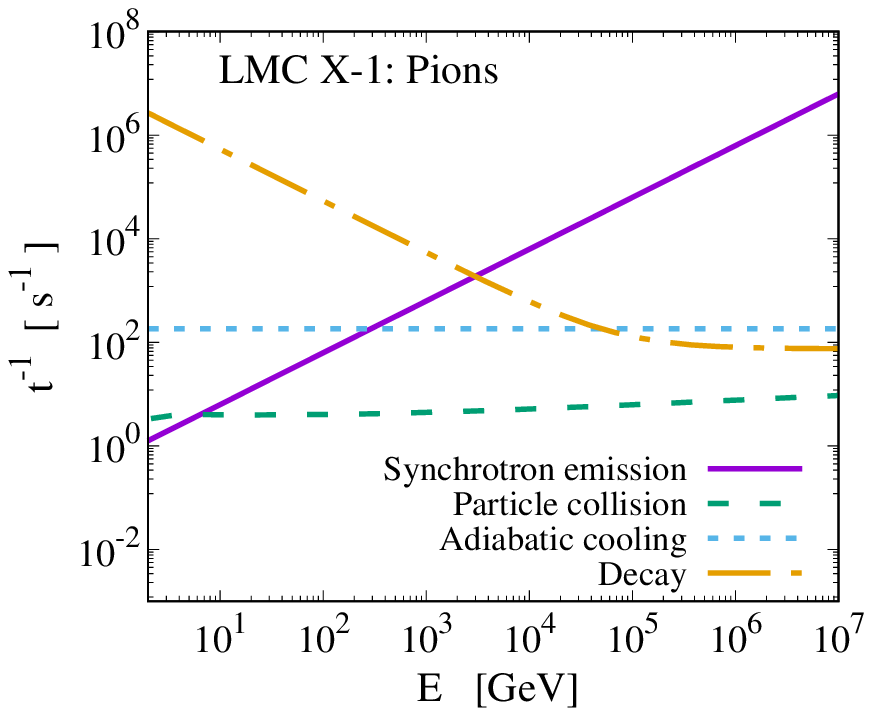}
\hspace*{-0.6 cm}
\includegraphics[width=.45\linewidth]{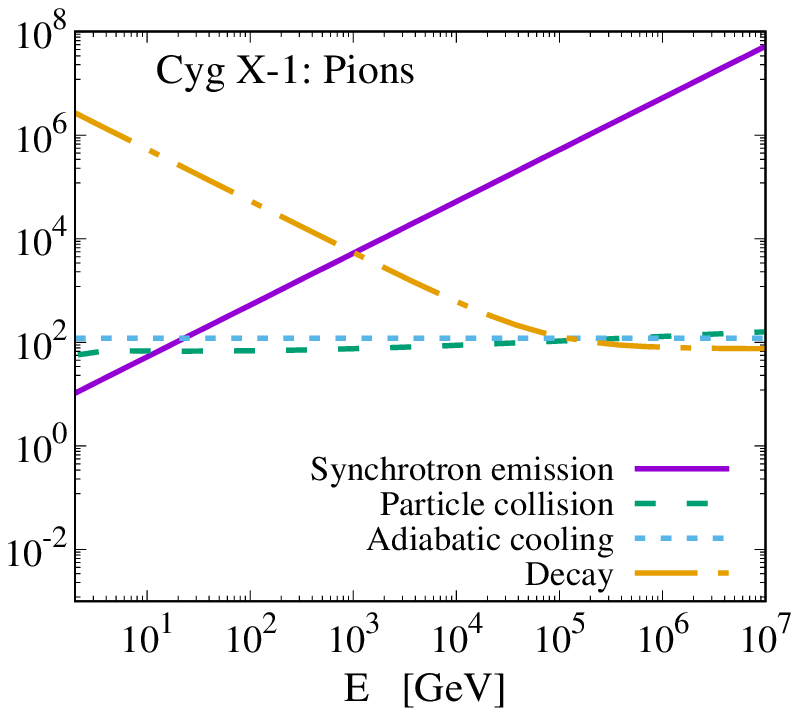}\\
\includegraphics[width=.45\linewidth]{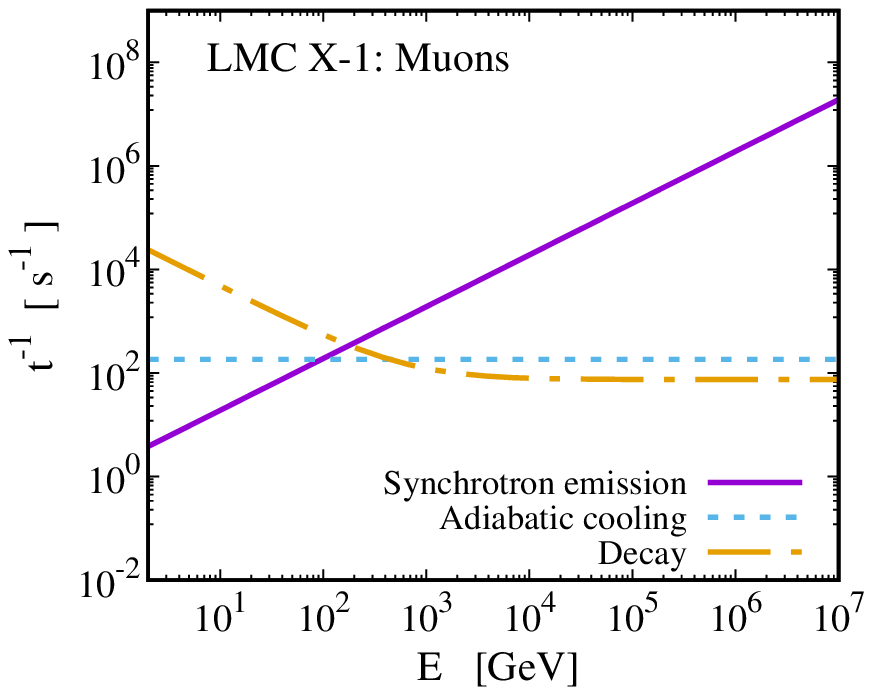}
\hspace*{-0.6 cm}
\includegraphics[width=.45\linewidth]{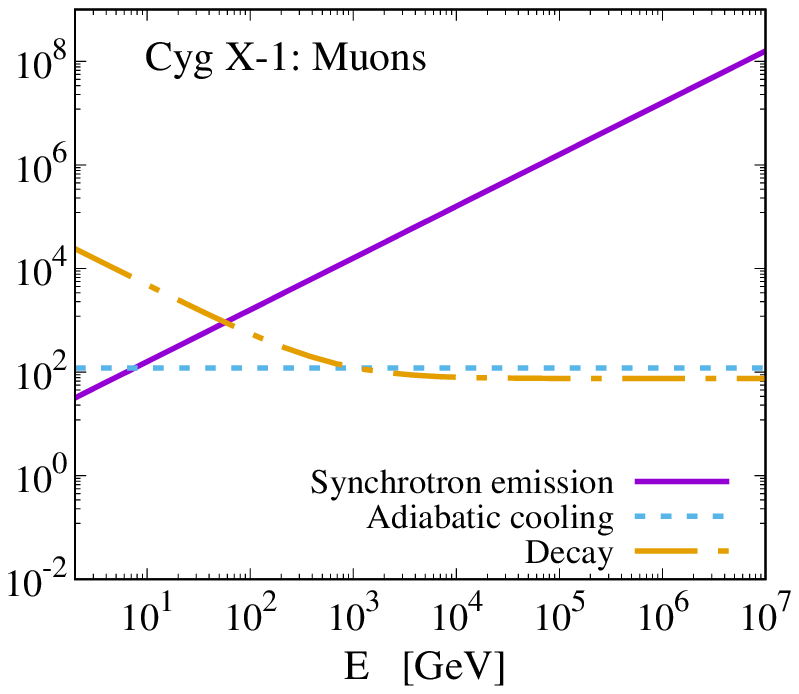}
\vspace*{0.2 cm}
\caption{\label{figure1} Cooling rates for the relativistic protons and 
the secondary particles ($\pi^\pm$ and $\mu^\pm$)
produced after the p-p collision processes take place in LMC X-1 (first 
column) and Cygnus X-1 (second column).}
\end{figure*}

All particles that take part in the neutrino and gamma-ray production processes lose 
energy while traveling along the acceleration zone, which could be due to different 
mechanisms as it is illustrated in Fig. \ref{figure1}. At first, the particles can be subjected to adiabatic 
energy losses due to jet expansion along the ejection axis with a rate depending on 
the jet's bulk velocity linearly \cite{Bosch-Ramon}. Another important cooling mechanism, that depends on the cold proton density inside the jet, is due to inelastic collisions of the accelerated
particles with the cold ones.  
The inelastic cross section for the p-p scattering is given by \cite{Kelner, Ody-Smpon-2015, Ody-Smpon-2017,Ody-Smpon-2018}
\begin{align}
\sigma _{pp}^{inel}(E_p)=(0.25L^2+1.88L+34.3) \nonumber \\ \times\left[1-
\left(\frac{E_{th}}{E_p}\right)^4\right]^2\times 10^{-27} cm^2
\end{align}      
where $L=ln(E_p/1000)$ with $E_p$ in GeV and $E_{th}=1.2$ $GeV$ the threshold for the production 
of a single neutral pion. Respectively, for pions it holds \cite{Gaisser}
\begin{equation}
\sigma _{\pi p}^{inel}(E_{\pi})\simeq \frac{2}{3}\sigma _{pp}^{inel}(E_{\pi})
\end{equation} 
In addition, particles accelerated by the magnetic fields emit synchrotron radiation. 
Thus, they gradually lose part of their energy with a rate that heavily depends on the magnetic field
and the particle energy. 

Furthermore, protons and pions interact with radiation fields coming from internal
or external regions of the jet, such as the system's accretion disc \cite{Dermer}, as well as the synchrotron-emitting particles themselves. This leads to partially energy loss. Smaller particles, such as muons, transfer part of their
energy to low-energy photons via inverse Compton scattering. However, such contributions 
can be ignored compared to those mentioned above. 

\section{Solution of the transfer equation} 

The steady-state transfer equation, which fulfills the conditions discussed, is given by \cite{Ody-Smpon-2015,Ody-Smpon-2017,Ody-Smpon-2018}
\begin{equation}
\frac{\partial N(E,z)b(E,z)}{\partial E}+t^{-1}N(E,z)=Q(E,z) \, ,
\label{tranf-equat}
\end{equation} 
where $N(E,z)$ denotes the particle energy distribution in units of $GeV^{-1}cm^{-3}$ while 
$Q(E,z)$ is the particle source (or injection) function giving the respective production rate 
(in $GeV^{-1}cm^{-3}s^{-1}$). Concerning the energy loss rate $b(E)$, all the cooling 
mechanisms discussed are being represented as it is evident from its definition $b(E)=dE/dt=-Et_{loss}^{-1}$. 

For each particle, the decay rate is added to the particle escape rate from the jet according to the relation: $t^{-1}=t_{esc}^{-1}+t_{dec}^{-1}$. The escape rate is approximated to $t_{esc}^{-1}=c/(z_{max}-z)$, 
where $z_{max}$ stands for the end of the acceleration zone.

The general solution of the differential equation (\ref{tranf-equat}) is given by 
\begin{equation}
N(E,z)=\frac{1}{\mid b(E) \mid}\int_{E}^{E_{max}} Q(E',z)e^{-\tau (E,E')}dE' \, .
\end{equation}
where $\tau (E,E')=\int_{E}^{E'} (dE"t^{-1})/\mid b(E")\mid$.
Therefore, in order to calculate the neutrino and gamma-ray emissivities, it is necessary to calculate first the energy distributions of all particles involved in the reaction chain (see sec. II).    

\subsection{Particle injection functions}

\subsubsection{Relativistic proton injection}

In previous works \cite{Papav-Papad-Kosm}, the appropriate injection function for the relativistic protons produced by the Fermi acceleration mechanism was found to be power-law with exponent $\approx 2$ \cite{Achterberg, Kirk}. In the jet's rest frame, this power-law translates to the following expression 
\begin{align}
Q(E',z)=Q_0\left(\frac{z_0}{z}\right)^3E'^{-2}
\label{Prot_sourc}
\end{align} 
where $Q_0$ is a normalization constant calculated through the total luminosity that is carried by the protons (or electrons) inside the jet (see Appendix \ref{Normal}), where $E_{p}^{min}=1.2$ $GeV$ is the minimum 
proton energy that is sufficient and necessary for the Fermi mechanism to occur. The maximum energy is calculated by equating the particle acceleration rate with the total energy loss rate $t_{acc}^{-1}\approx \eta ceB/E_p=t_{loss}^{-1}$. For the binary systems of our interest, this is approximated as $E_{p}^{max}\simeq 10^7$ $GeV$. The dependence on z is due to particle conservation enforced on the respective current density \cite{Ghisellini}. Transformation of the source function of Eq. (\ref{Prot_sourc}) to the observer's 
reference frame givens \cite{Ody-Smpon-2017,Ody-Smpon-2018}
\begin{align}
Q(E,z)=\frac{Q_0(\frac{z_0}{z})^3}{\Gamma_b(E-\beta_b cos\theta\sqrt{E^2-m^2c^4})^2}\nonumber \\ \times\left(1-\frac{\beta_bEcos\theta}{\sqrt{E^2-m^2c^4}}\right)
\end{align}
where $\Gamma_b$ responds to the jet's Lorentz factor and $\theta$ is the angle between the jet's ejection axis and the line of sight.  

\subsubsection{Pion energy distribution}

The pion source function calculation requires the fast proton distribution as well as the p-p collision rate along with the pion spectra produced by each one of those collisions as
\begin{align}
Q_{\pi}(E,z)=cn(z)\int_{\frac{E}{E_{max}}}^{1}N_p\left(\frac{E}{x},z\right)F_{\pi}\left(x,\frac{E}{x}\right)\nonumber \\ \times\sigma_{pp}^{inel}\left(\frac{E}{x}\right)\frac{dx}{x}
\label{Pion_distr}
\end{align} 
where $x = E/E_p$. In the latter integral, $F_\pi (x,E/x)$ denotes the pion mean number produced per p-p 
collision given by \cite{Kelner}
\begin{align}
F_{\pi}\left(x,\frac{E}{x}\right)=4\alpha B_{\pi}x^{\alpha -1}\left(\frac{1-x^{\alpha}}{1+rx^{\alpha}(1-x^{\alpha})}\right)^4 \nonumber \\ \times\left(\frac{1}{1-x^{\alpha}}+\frac{r(1-2x^{\alpha})}{1+rx^{\alpha}(1-x^{\alpha})}\right)\left(1-\frac{m_{\pi}c^2}{xE_p}\right)^{\frac{1}{2}}
\end{align}
where $B_{\pi}=\alpha '+0.25$, $\alpha '=3.67+0.83L+0.075L^2$, $r=2.6/\sqrt{\alpha '}$ and $\alpha=0.98/\sqrt{\alpha '}$. Also, $n(z)$ is the cold proton density of the jet written as
\begin{equation}
n(z) = \frac{(1-q_r)L_k}{\Gamma m_pc^2\pi r(z)^2\upsilon _{b}}
\end{equation}
where $\Gamma$ is the cold proton Lorentz factor. From Eq. (\ref{Pion_distr}), it is worth noting that the proton 
distribution is entering both the neutrino and gamma-ray emissivity calculations through the pion injection rate $Q_{\pi}(E,z)$.

\begin{figure*}[ht] 
\centering
\includegraphics[width=.45\linewidth]{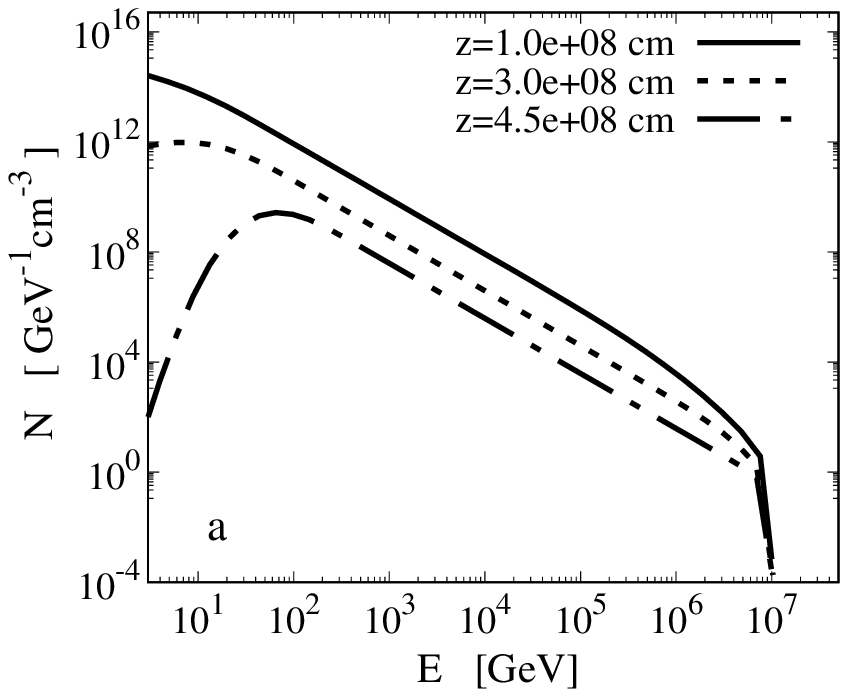}
\hspace*{-0.6 cm}
\includegraphics[width=.45\linewidth]{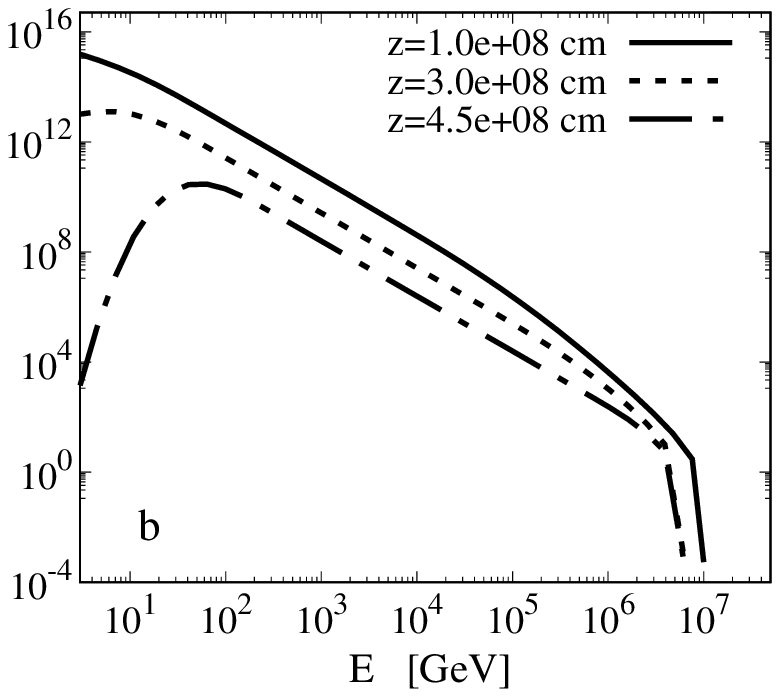}\\
\includegraphics[width=.45\linewidth]{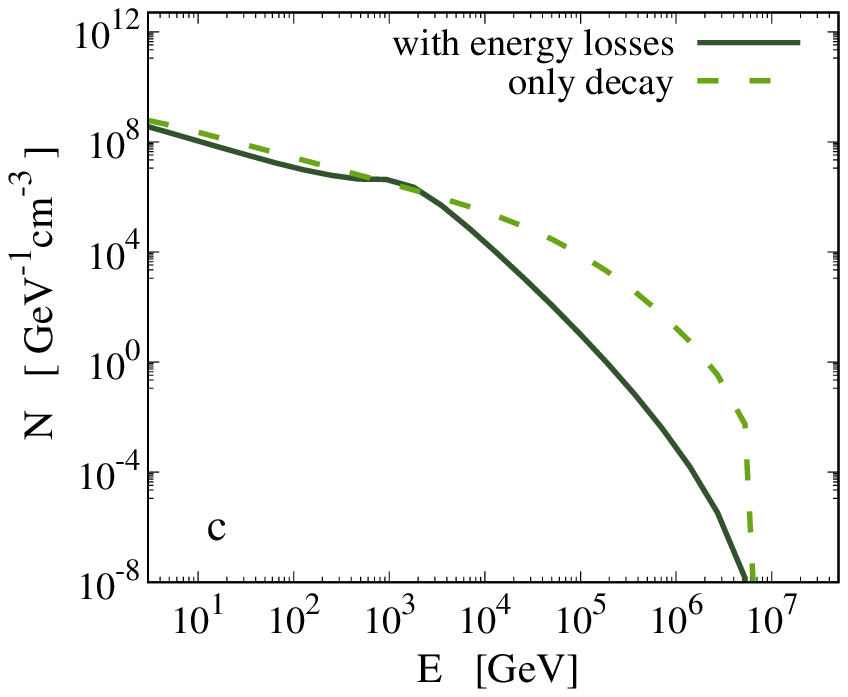}
\hspace*{-0.6 cm}
\includegraphics[width=.45\linewidth]{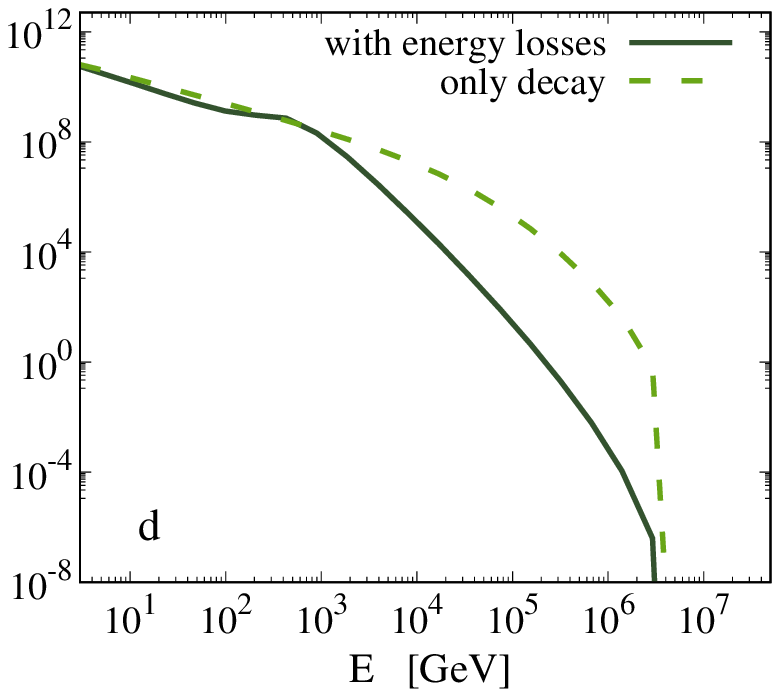}\\
\includegraphics[width=.45\linewidth]{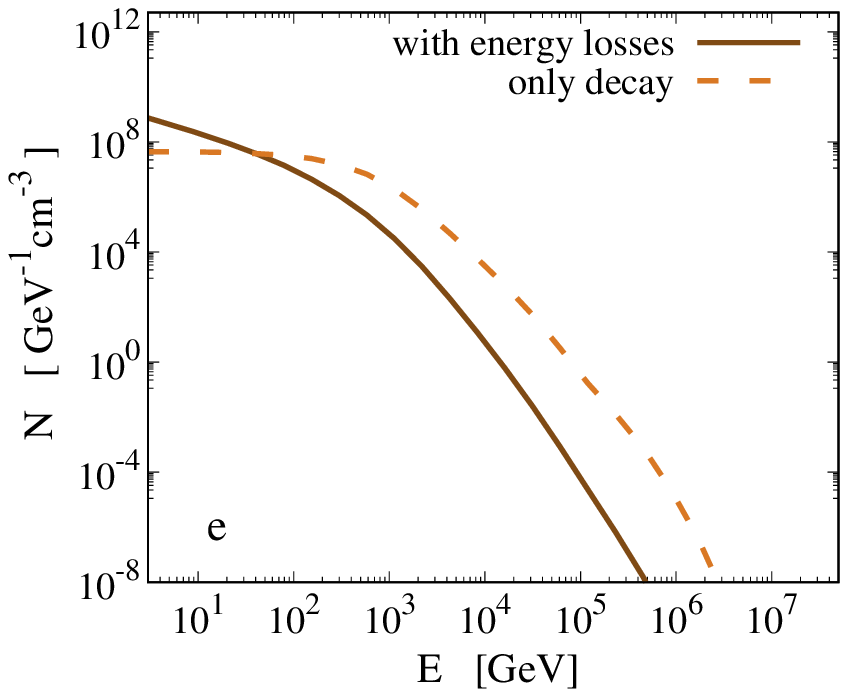}
\hspace*{-0.6 cm}
\includegraphics[width=.45\linewidth]{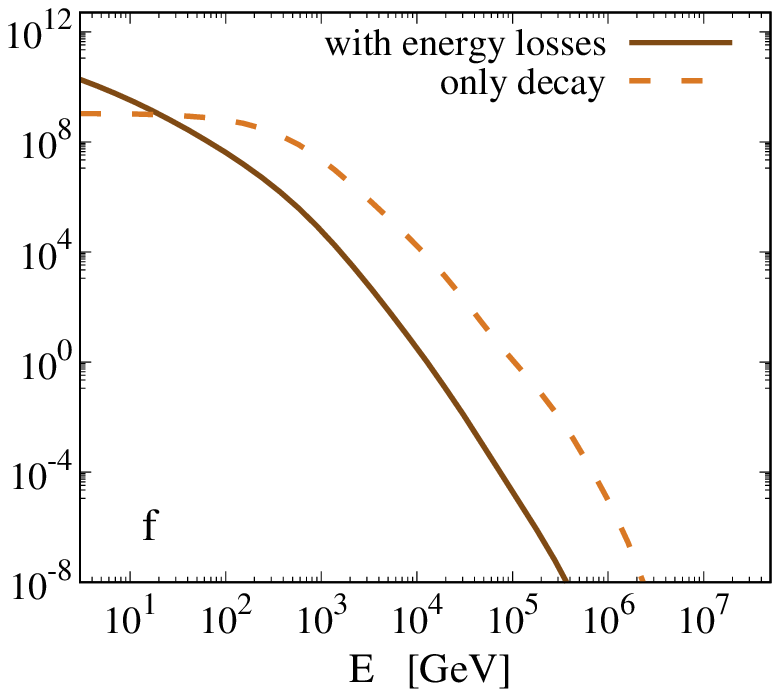}
\vspace*{0.2 cm}
\caption{\label{figure2} Energy distributions for the relativistic protons (a, b), pions ($\pi^\pm$) (c, d) and muons ($\mu^\pm$) (e, f) produced in LMC X-1 (a, c, e) and Cygnus X-1 (b, d, f).}
\end{figure*}

\subsubsection{Muon spectra from pion decay}

As it is known, for the muon energy distribution, both the mean right-handed and the mean left-handed muon numbers per pion decay are required for obtaining the total injection function. According to the CP invariance and provided 
that $N_{\pi}(E_{\pi},z)=N_{\pi ^{+}}(E_{\pi},z)+N_{\pi ^{-}}(E_{\pi},z)$, it holds 
\cite{Lipari} 
\begin{align}
Q_{\mu _R^{\pm},\mu _L^{\mp}}(E_{\mu}, z)=\int_{E_{\mu}}^{E_{max}} 
dE_{\pi}t_{\pi ,dec}^{-1}(E_{\pi})N_{\pi}(E_{\pi},z)\nonumber \\ \times\mathcal{N} _{\mu}^{\pm}\Theta (x-r_{\pi})
\label{Muon_distr}
\end{align}
where $\mathcal{N} _{\mu}^{+}$ and $\mathcal{N} _{\mu}^{-}$ represent the positive and negative right (or inversely the left) handed muon spectra, respectively (see Appendix \ref{muon_spectra}).
In Eq. (\ref{Muon_distr}), we have used $x = E_\mu/E_\pi$, $r_\pi =(m_\mu/m_\pi)^2$ 
and $\Theta (y)$ the Heaviside function. Additionally, the pion decay rate is given by 
$t_{\pi ,dec}^{-1}=(2.6\times 10^{-8}\gamma _{\pi})^{-1}$ $s^{-1}$, which implies
that the pion distribution is important for the muon distribution calculation.

\section{Results and discussion}

\subsection{Neutrino spectra from pion and muon decay}

After obtaining the energy spectra of the particles discussed above, one is able to estimate the total number of neutrinos produced directly from pion decay as well as from muon decay. Thus, 
the total emissivity contains both contributions as
\begin{equation}
Q_{\nu}(E,z)=Q_{\pi\rightarrow\nu}(E,z)+Q_{\mu\rightarrow\nu}(E,z)
\end{equation}  
The first term represents the neutrino injection originating from pion decay as
\begin{align}
Q_{\pi\rightarrow\nu}(E,z)=\int_E^{E_{max}}t_{\pi,dec}^{-1}(E_\pi) N_\pi(E_\pi,z)\nonumber \\ \times\frac{\Theta 
(1-r_\pi-x)}{E_\pi(1-r_\pi)} dE_\pi
\end{align} 
where $x=E/E_{\pi}$, while the second term gives
\begin{align}
Q_{\mu\rightarrow\nu}(E,z)=\sum_{i=1}^4\int_E^{E_{max}}t_{\mu,dec}^{-1}(E_\mu)N_{\mu_i}(E_\mu,z)  
\nonumber \\ \times\left[\frac{5}{3}-3x^2+\frac{4}{3}x^3+(3x^2-\frac{1}{3}-\frac{8}{3}x^3)h_i\right]\frac{dE_\mu}{E_\mu}
\end{align}
with $x=E/E_\mu$. In the latter equation, the muon decay rate depends on their energy as follows $t_{\mu ,dec}^{-1}=(2.2\times 10^{-6}\gamma_\mu)^{-1}$ $s^{-1}$. 
Also $h_3=h_4=-h_1=-h_2=1$. From the four different integrals of the latter summation, the first 
and second represent the left-handed muons of positive and negative charge, respectively, while 
the third and fourth stand for the corresponding right-handed ones. 
Finally, integration over the acceleration zone gives the total neutrino intensity as \cite{Reynoso,Romero}
\begin{align}
I_{\nu}(E)&=\int_VQ_{\nu}(E,z)d^3r\nonumber \\ &=\pi (tan\xi)^2\int_{z_0}^{z_{max}} Q_{\nu}(E,z)z^2dz
\end{align}

\subsection{Gamma-ray emissivity for $E>100$ GeV}

In this work, we assumed that gamma-ray production is mainly due to neutral pions decay which in turn are products of p-p inelastic collisions. The respective gamma-ray spectra have been simulated for photons of energy $E_{\gamma}=x{E_p}$ to be \cite{Kelner} 
\begin{align}
F_{\gamma}(x,E_p)=B_{\gamma}\frac{lnx}{x}\left(\frac{1-x^{\beta_{\gamma}}}{1+k_{\gamma}x^{\beta_{\gamma}}(1-x^{\beta_{\gamma}}}\right)^4\nonumber \\ \times\left(\frac{1}{lnx}-\frac{4\beta_{\gamma}x^{\beta_{\gamma}}}{1-x^{\beta_{\gamma}}}-\frac{4k_{\gamma}\beta_{\gamma}x^{\beta_{\gamma}}(1-2x^{\beta_{\gamma}})}{1+k_{\gamma}x^{\beta_{\gamma}}(1-x^{\beta_{\gamma}})}\right)
\label{Gamma-ray spectra}
\end{align}
where $B_{\gamma}=1.3+0.14L+0.011L^2$, $\beta_{\gamma}=1/(0.008L^2+0.11L+1.79)$ and $k_{\gamma}=1/(0.014L^2+0.049L+0.801)$. In addition, we have $L=ln(E_p/1 TeV)$. These results are consistent with proton energies in the range $0.1$ $TeV<E_p<10^5$ $TeV$. Besides $\pi ^0$, Eq. (\ref{Gamma-ray spectra}) also considers the contribution of $\eta$ mesons decay, which is approximately 25\% when $x\approx 0.1$.\\
  
 In the energy range of our interest $E_{\gamma}\geq 100$ $GeV$, the gamma-ray emissivity, produced at a distance $z$ from the compact object along the jet's ejection axis, is given by
\begin{align}
Q_{\gamma}(E_{\gamma},z)=cn(z)\int_{\frac{E_{\gamma}}{E_p^{max}}}^{1}\frac{dx}{x}N_p\left(\frac{E_{\gamma}}{x},z\right)\nonumber \\ \times F_{\gamma}\left(x,\frac{E_{\gamma}}{x}\right)\sigma _{pp}^{(inel)}\left(\frac{E_{\gamma}}{x}\right)
\end{align}            
(in units of $GeV^{-1}cm^{-3}s^{-1}$). The respective intensity is obtained again through integration over the acceleration zone. For $E_{\gamma}<100$ GeV, the delta-function approximation is employed.

\subsection{Neutrino and gamma-ray intensity simulations}  

By introducing the calculated rates for the cooling mechanisms of particles that lead to the neutrino and gamma-ray production, to the transfer equation discussed previously along with the corresponding injection function, we are able to calculate the particle energy distributions. These results are shown in Fig. \ref{figure2}. The reaction chain seen above includes the relativistic protons accelerated by shock-waves through the Fermi mechanism, the charged pions produced by the p-p interactions and the muons that result from the pion decays. We consider all the above mechanisms and interactions taking place in the hadronic relativistic jets of the extragalactic binary system LMC X-1 located in our galaxy-neighbor LMC (Large Magellanic Cloud) and with a distance of 55 kpc to the Earth \cite{diBenedetto}. For comparison, the corresponding results for the galactic Cygnus X-1 are also shown. Then, the energy spectra of the produced high-energy neutrinos and gamma-rays can be simulated numerically by using those calculations.  

The calculations were performed through the development of a C-code (that employs Gauss-Legendre numerical integration of the GSL library) and the use of the parameter values listed in Table \ref{Table1} mostly describing geometric characteristics of the systems of interest.


\begin{table}
\caption{\label{Table1} Parameters describing geometric characteristics of the extragalactic 
LMC X-1, in the Large Magellanic Cloud, and the Galactic Cygnus X-1 binary systems.}
\begin{center}
\begin{tabular}{l l l l}
\hline \\ [0.01ex] 
Description & Parameter & LMC X-1 & Cyg X-1 \\
\hline \\ [0.01ex]
Jet's base & $z_0$ & $1\times 10^8$ $[cm]$ & $1\times 10^8$ $[cm]$ \\ [0.2ex]
Acceleration limit & $z_{max}$ & $5\times 10^8$ $[cm]$ & $5\times 10^8$ $[cm]$ \\ [0.2ex]
Black Hole mass & $M_{BH}$ & 10.91\(M_\odot\)\cite{Orosz2009} & 14.8\(M_\odot\)
\cite{Orosz2011} \\ [0.2ex]
Angle to the line-of-sight & $\theta$ & 36.38$^\circ$\cite{Orosz2009} & 27.1$^\circ$
\cite{Orosz2011} \\ [0.2ex]
Jet's half-opening angle & $\xi$ & 3$^\circ$ & 1.5$^\circ$ \\ [0.2ex]
Jet's bulk velocity & $\upsilon _{b}$ & 0.92c & 0.6c \cite{Stirling} \\ [0.2ex]
\hline
\end{tabular}
\end{center}
\end{table} 


At first, we calculated the fast proton distributions for three different distances to the central object for binary systems LMC X-1 and Cygnus X-1 as illustrated in the graphs a and b, respectively, of Fig. \ref{figure2}. It is evident that the total particle density production decreases while we move closer to the end of the acceleration zone even though the average particle energy increases.

In Fig. \ref{figure2} graphs c and d, we show the energy distributions for pions that have suffered energy losses caused by various mechanisms such as synchrotron radiation emission, collisions with the rest of the jet matter etc. \cite{Papav-Papad-Kosm}. For comparison, we also plot the respective distributions of particles that do not lose energy at all. This exhibits the important effect which the cooling mechanisms cause on the total particle distributions. The same can be concluded for the muon distributions from the graphs e and f in Fig. \ref{figure2}. We also notice that the cooling rate characteristics are reflected upon the particle distributions as can be seen from the deviation of the two lines in the pion and muon case. Mainly, the deviation point coincides with the dominance of the synchrotron mechanism over the particle decay. The synchrotron losses take off causing the smoother transition in the solid line's case (with energy losses) compared to the dashed one (only decay)\cite{Papav-Papad-Kosm}.   

\begin{figure}[ht]
\centering
\includegraphics[width=\linewidth]{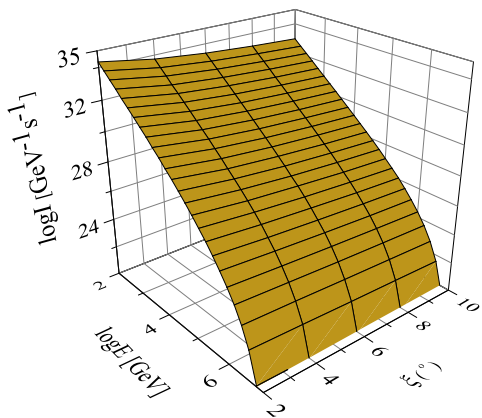}
\caption{\label{figure3} 3-D graph of the gamma-ray intensity produced in LMC X-1 presents its dependence on the jet's half-opening angle $\xi$ along with the energy.}
\end{figure}

Furthermore, an instant observation would be that there is not significant difference between the two binary systems.
This is a result of their black hole masses being of similar magnitude. The parameter that has the biggest impact though is the jet's half-opening angle, which roughly gives the degree of the jet's collimation. That is because the system's magnetic field, which is responsible for the collimation, presents a strong dependence on $\xi$.   

\begin{figure*}[ht] 
\centering
\includegraphics[width=.5\linewidth]{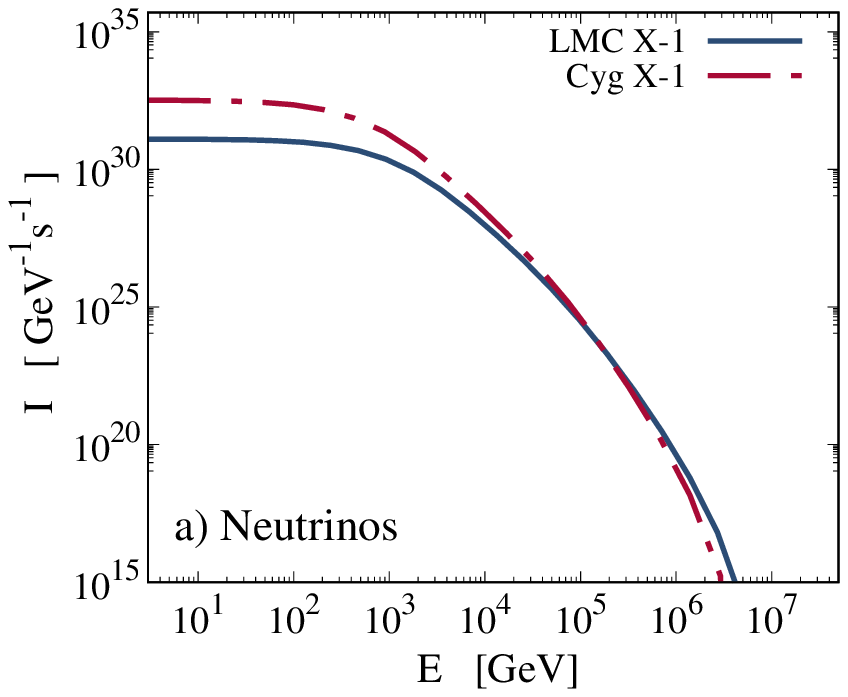}
\hspace*{-0.7 cm}
\includegraphics[width=.5\linewidth]{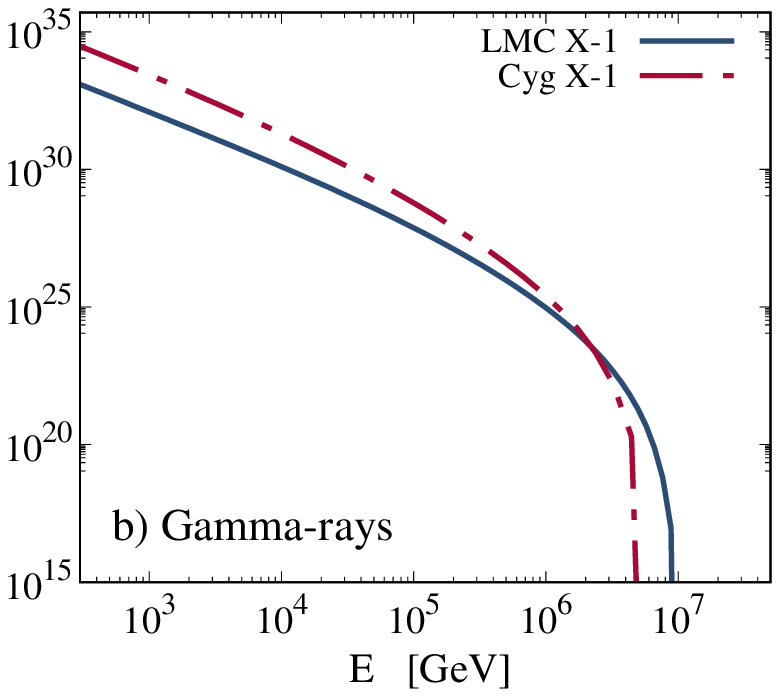} \\
\vspace*{0.2 cm}
\caption{\label{figure4} Neutrino (a) and gamma-ray (b) intensities produced by pion ($\pi ^{\pm}$) decay that in turn are products of the p-p interaction mechanism taking place in the extragalactic binary system LMC X-1 as well as the well-studied system Cygnus X-1.}
\end{figure*}

After calculating all the necessary particle distributions, the neutrino and gamma-ray 
emissivities as well as the corresponding intensities are readily obtained in Fig. \ref{figure3} and \ref{figure4}. 
For the LMC X-1 system, our results have shown 
that, the increase of the half-opening angle $\xi$ leads to a decrease in the 
gamma-ray production, which is an expected result since the p-p collision 
rate drops with the jet's expansion as can be seen in Fig. \ref{figure3}.

\section{Summary and Conclusions}

During the last decades, the structure and evolution of relativistic astrophysical plasma outflows (jets) and specifically those connected  
to compact cosmic structures, became research subjects of intense interest. Towards this aim, a great research effort, experimental, theoretical, 
and phenomenological is absorbed by various high-energy phenomena, including production and terrestrial detection of high-energy cosmic 
radiation and neutrinos, originating from Galactic and extragalactic sources heading towards the Earth. We mention in particular, the 
jets of very large scale ejected from galaxy's quasar systems involving supermassive black holes at the central region (AGN) and those
of much smaller scale involving stellar mass black holes and a companion star, known as micro-quasars and X-ray binary systems.

In this work, we concentrated on the latter class, the two-body systems consisting of a central object (usually a stellar mass black hole) and 
a companion star often of O-type or B-type main sequence stars. The former absorbs mass from the latter, forming an accretion disc of gas and 
matter which emits X-ray radiation due to very high temperatures prevailing in the area of accretion disk and black hole. The jets are formed 
when the system's magnetic field collects matter that is ejected away from the system in a collimated and accelerated bulk-like plasma flow. 

The mass outflow can acquire relativistic velocities and is expelled perpendicular to the disc's surface. In many models these jets are
treated magnetohydrodynamically by assuming various reliable approximations. In the present work, we considered the jet's matter to be mainly 
hadronic, with a portion of it accelerated through shock-waves to relativistic velocities. From the inelastic collision of relativistic 
protons on the cold ones (p-p interactions mechanism), secondary neutral and charged particles (pions, kaons, muons, etc.) are produced,
the decay of which leads to high-energy neutrino and gamma-ray emissions. 
 
One of our main goals was the calculation of the energy-spectra of high energy neutrino and gamma-ray produced inside such astrophysical
jets. As concrete examples, we have chosen the Galactic X-ray binary Cygnus X-1 system and the extragalactic LMC X-1 binary in order to
simulate their neutrino and gamma-ray intensities emitted. For the observation of such high energy cosmic radiations and particle (neutrino) 
emissions, extremely sensitive detection instruments are operating and next generation detectors have been designed at the Earth like the
IceCube detector (deep under the ice at South Pole), the ANTARES and KM3NeT (underwater in the Mediterranean sea), the CTA etc.   

\section*{Acknowledgments} This research is co-financed (O.T.K) by Greece and the European Union (European Social Fund-ESF) through the Operational Programme "Human Resources Development, Education and Lifelong Learning 2014-2020" in the context of the project (MIS5047635). Th.V.P wishes to thank Prof. T.S. Kosmas for fruitful discussions during my stay in the Dept. of Physics, University of Ioannina.

\section*{Data Availability}

There is no data used in this paper.

\section*{Conflicts of Interest} The authors declare that there are no conflicts of interest regarding the publication of this paper.

\appendix

\section{}

\subsection{Normalization constant}

\label{Normal}
The normalization constant for the relativistic proton source function depends on the lower and upper limit of their energy as follows
\begin{align}
Q_0=\frac{8q_rL_k}{z_0r_0^2ln(E_p^{max}/E_p^{min})}
\end{align}
The above result is calculated through the total luminocity carried by the protons that is given by 
\begin{align}
L_{p}=\int_{V}d^3r\int_{E_{p}^{min}}^{E_{p}^{max}}E_{p}Q_{p}(E_{p},z)dE_{p}
\end{align}
\\
\subsection{Right and left-handed muon spectra}

\label{muon_spectra}
The right-handed positive and negative muon spectra produced by pion decay are
\begin{equation}
\mathcal{N} _{\mu}^{+}=\frac{r_{\pi}(1-x)}{E_{\pi}x(1-r_{\pi})^2} \, , \quad 
\mathcal{N} _{\mu}^{-}=\frac{(x-r_{\pi})}{E_{\pi}x(1-r_{\pi})^2}
\end{equation}
where $x = E_\mu/E_\pi$ and $r_\pi =(m_\mu/m_\pi)^2$. According to CP invariance, the number of $\mu _{L}^{-}$ produced by the $\pi ^{-}$ decay is the same as the $\mu _{R}^{+}$ produced by the $\pi ^{+}$ decay. Therefore, as the pion energy distribution refers to the sum of both $\pi ^{+}$ and $\pi ^{-}$ distributions, it is $\mathcal{N} _{\mu ^{R}}^{+}=\mathcal{N} _{\mu ^{L}}^{-}$. Similarly, it holds the same for the negative right-handed muons.

\end{document}